\documentclass[aps, pra, amsmath, preprintnumbers, superscriptaddress, twocolumn, amssymb, floatfix]{revtex4-1}
\usepackage{graphicx}
\usepackage{mathptmx, textcomp}
\usepackage[latin1]{inputenc}

\begin{document}
\title{Cold atom-ion experiments in hybrid traps}
\author{Arne H\"arter}
\affiliation{Institut f\"ur Quantenmaterie and Center for Integrated Quantum Science and Technology IQ$^\textrm{ST}$,
Universit\"at Ulm, 89069 Ulm, Germany}
 
\author{Johannes Hecker Denschlag}
\affiliation{Institut f\"ur Quantenmaterie and Center for Integrated Quantum Science and Technology IQ$^\textrm{ST}$,
Universit\"at Ulm, 89069 Ulm, Germany}
             
\begin{abstract}
In the last 5 years, a novel field of physics and chemistry has
developed in which cold trapped ions and ultracold atomic gases are
brought into contact with each other. Combining ion traps
with traps for neutral atoms yields a variety of new possibilities
for research and experiments. These range from studies of cold
atom-ion collisions and atom-ion chemistry to applications in
quantum information science and condensed matter related research. In this
article we give a brief introduction into this new field and describe some of the
perspectives for its future development.
\end{abstract}

\maketitle

\section{Introduction}
The two research fields of trapped ultracold atoms and trapped
laser-cooled ions have undergone a spectacular development during
the last decades. They can be viewed as two cornerstones of
today's research on quantum physics and they are continuing to
push the limits of this
field$\,$\cite{Lei03,Bla08,Haf08,BDZ2008,Blo12}. Only quite recently,
however, have
trapping technologies for ions and neutral atoms
been combined in hybrid setups to allow for the controlled
immersion of ions into clouds of ultracold
atoms$\,$\cite{Gri09,Smi2010,Zip10}.

Traps for ions and traps for neutral atoms have complementary
properties that turn hybrid atom-ion setups into versatile and powerful
experimental tools.
On the one hand, Paul traps for ions can be tight and several
eV deep, allowing for a strong localization of the ions and for almost
arbitrarily long storage times.
On the other hand, even in such tight traps, laser-cooled ions are typically
separated spatially by several micrometers due to their electrostatic
repulsion. Thus, they can be individually optically detected and addressed.
A single ion can then be employed as a well localized probe, e.g., to investigate
local properties of an atomic cloud. In contrast,
neutral atoms do not interact nearly as strongly as their charged
counterparts. They can thus be prepared in large, dense ensembles
consisting of millions of particles. Evaporative cooling methods
provide access to temperatures on the nano-kelvin scale and to
Bose-Einstein condensation. Ultracold clouds of atoms therefore
lend themselves to sympathetic cooling of ions. Furthermore,
they can also be used as sensitive bolometers
to measure energy deposited by single ions, as 
even small atomic temperature changes can be detected. These first examples already
give an impression of how the atomic and ionic systems complement each other
and how they combine to form a fascinating experimental playground.

In this article we provide an introduction into
the field of cold atom-ion experiments. We intend to give the
reader a feeling for the characteristics of atom-ion interactions,
highlight the most important experimental factors and give a
rough overview of the current status of research activities. Finally,
we also briefly discuss some possible future directions of the field.
The discussions within this review should not be considered a complete
summary of the field and the reader is encouraged to further reading
starting with the references given here.
This article is organized as follows: In section 2 we study the
${1}/{r^4}$ interaction potential between an ion and an atom. It
is much more long-range than the $1/r^6$ van-der-Waals potential
between neutral ground state atoms, but of shorter range than the
$1/r^3$ interaction between dipoles. In section 3 we give an
overview of typical experimental realizations of atom-ion hybrid
traps. While atom and ion traps are in principle well-compatible,
ion micromotion in the Paul trap represents an experimental
obstacle for atom-ion experiments in the lowest energy regime. As
discussed in section 4 in the context of elastic atom-ion collisions,
this trap-driven micromotion leads to
heating effects and sets a lower limit to the attainable collision
energies. In section 5 we then turn to cold chemical reactions
between atoms and ions, which is a vast field of research. Here,
we constrain ourselves to the recent developments
with respect to cold reactions in hybrid traps.
One important reaction is charge transfer where the positive
charge of the ion is passed onto an atom of the neutral atom gas.
Furthermore, we also touch upon quenching collisions and
the formation of charged molecules. Finally, in section 6 we review
some future prospects for atom-ion experiments. On the one hand,
we cover ideas and first experiments in the direction of
quantum information science. On the other hand, prospects in the context
of condensed matter physics are discussed. In particular,
we consider charge transport in a quantum gas, impurity physics, and possibilities of
emulating solid state physics.

\section{The atom-ion interaction}
\label{sec:potential}

The atom-ion interaction potential is dominated by the effect of
the electric field $\varepsilon(r)=e/(4\pi\epsilon_\text{0}r^2)$
of the ion on the polarizable atom. Here, $e$ is the elementary charge,
$\epsilon_\text{0}$ the vacuum permittivity and $r$ the separation between
atom and ion. The electric field induces a dipole moment $p(r)=4\pi
\epsilon_\text{0} \alpha \varepsilon (r)$ in the atom which has a
polarizability $\alpha$. Due to the inhomogeneity of the
field, the induced dipole feels an attractive force towards the
ion.
\begin{figure}
\begin{center}
\resizebox*{7cm}{!}{\includegraphics{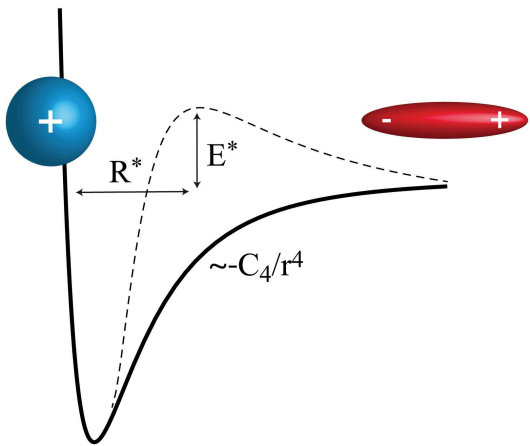}}%
\caption{The atom-ion polarization potential. The centrifugal
barrier of the p-wave scattering channel (dashed line) leads to an
energy scale $E^*$ and a length scale $R^*$.}%
\label{fig:potential}
\end{center}
\end{figure}
As a result, the atom-ion interaction is described by the
polarization potential
\begin{equation}
V(r)=-\frac{1}{2}p(r)\varepsilon (r)=-\frac{C_4}{r^4}
\end{equation}
with $C_4=\alpha e^2/(8\pi\epsilon_\text{0})$. In addition to this
scattering potential, the angular momentum potential $l(l+1) \hbar^2 / 2\mu r^2 $
has to be considered, where  $l$ is the angular momentum quantum number  
and $\mu$ is the reduced mass of the atom-ion system. As a consequence, for p-wave
scattering ($l = 1$) the resulting effective potential exhibits a barrier
(see Fig.$\,$\ref{fig:potential}). It defines two useful scales,
 namely a length scale $R^*=\sqrt{2 \mu C_4/\hbar^2}$ and an energy scale $E^*=\hbar^2/(2\mu R^{*2})$.
The length
scale $R^*$ determines an interaction range and sets a typical
value for the scattering length $a$. For the example of a
$^{138}$Ba$^+$ ion colliding with a $^{87}$Rb atom, $R^*$ amounts
to $R^*_\text{BaRb}=295\:$nm. This is about two orders of
magnitude larger than the typical range (or scattering length) for
van der Waals potentials between neutral atoms. The characteristic
energy $E^*$ indicates the onset of the s-wave regime. For Ba$^+$
and Rb we get $E^*_\text{BaRb}=50\:k_\text{B}\times$nK, which is
about three orders of magnitude lower as compared to the neutral
case. As we will explain in more detail later in the text, typical
atom-ion collision energies are currently technically limited to
values on the order of mK$\times k_\text{B}$ or higher. Thus,
reaching the s-wave scattering regime which would allow for control of atom-ion interactions
remains one of the challenges of the field.

\begin{figure}
\begin{center}
\resizebox*{8cm}{!}{\includegraphics{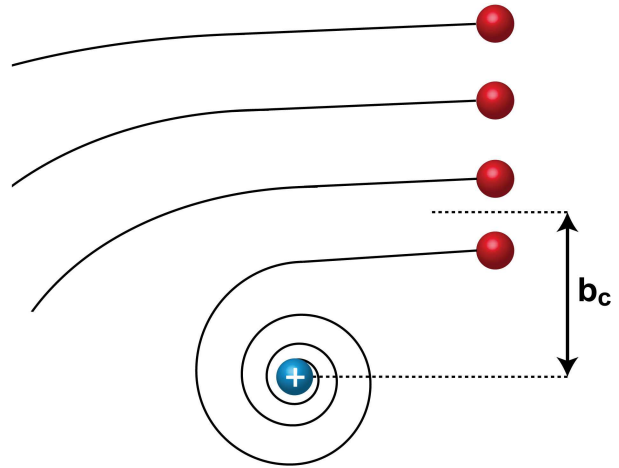}}%
\caption{Illustration of atom-ion collisions. Most collisions only
lead to weak deflection of the collision partners in so-called
glancing collisions. However, at impact parameters below the
critical value $b_c$, the attraction between atom and ion becomes
so strong, that the collision partners enter into an
inward-spiraling trajectory. This results
in a hard collision with a close encounter of the two nuclei.}%
\label{fig:spirale}
\end{center}
\end{figure}

To describe elastic atom-ion collisions in the currently
accessible higher energy regime one needs to account for
contributions of many partial waves. A corresponding semiclassical
expression for the elastic scattering cross section
$\sigma_\text{e}$ is given by $\,$\cite{Cot00}
\begin{equation}
\sigma_\text{e}\approx 5.08 \left( \frac{\mu
C_4^2}{\hbar^2}\right)^{1/3} E_\text{col}^{-1/3}\,,
\label{eq:sigmacote}
\end{equation}
where $E_\text{col}$ denotes the collision energy. Thus, the
semiclassical atom-ion cross section is energy-dependent and
increases with decreasing energy. Elastic atom-ion collisions can
be roughly grouped into two categories. 1) The glancing collisions
in which particles are only slightly deflected and are preferably
scattered in forward directions (see Fig.$\,$\ref{fig:spirale}).
These collisions are prevalent$\,$\cite{Zha09, Zip10b} and they
are observable e.g. in ion-induced heating effects in ultracold
atomic clouds.
 2) The second category are the Langevin
collisions in which two particles undergo a ``hard'' collision
where large momentum/ energy transfer is possible.  The Langevin
collisions can be viewed in a simple classical picture (see
Fig.$\,$\ref{fig:spirale}).
We consider an atom-ion  collision with an impact parameter $b$.
If the impact parameter is smaller than a critical value $b_c = (2
C_4 / E_\text{col})^{1/4}$, the two collision partners fall onto
each other in a spiraling motion. As already stated, significant
amounts of energy and momentum can be transferred between the
collision partners and the
direction into which the particles are scattered is isotropic. In
such a Langevin collision a reaction can also occur, e.g. in which
the charge is transferred from the ion to the atom or a molecule
is formed. In fact, as reactions of this sort usually do require a
close encounter of the two nuclei they will not occur in a
glancing collision. The Langevin cross section is given by
\cite{Lan05,Cot00}
\begin{equation}
\sigma_\text{Lgvn}=  \pi\sqrt{2C_4}E_\text{col}^{-1/2}\,.
\label{eq:ct}
\end{equation}
Considering a collision between Rb and Ba$^+$  at an energy of
about 1mK$\times k_B$, the Langevin cross section
$\sigma_\text{Lgvn}$ is about 10 times smaller than $
\sigma_\text{e}$. A detailed introduction into the theory of
atom-ion collisions, extending into the ultracold domain
of s-wave scattering, can be found in refs.$\,$\cite{Idz09} and \cite{Idz11}.\\

\section{Experimental realizations and their implications}
\label{sec:experimental}

\begin{figure*}
\begin{center}
\resizebox*{15cm}{!}{\includegraphics{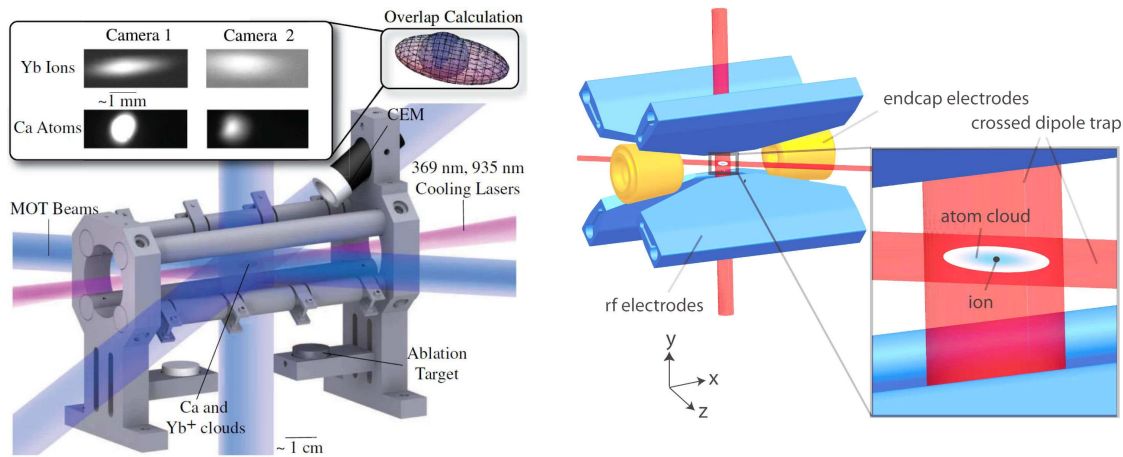}}%
\caption{Examples of hybrid atom-ion traps. \textit{Left}: A combined MOT and ion trap featuring large
separations of the Paul trap rf electrodes to allow for sizable MOT cooling beams inside the
trapping volume of the Paul trap. \textit{Right}: Alternative approach employing a crossed dipole trap
for atomic confinement and a significantly smaller spacing of the rf electrodes. Images taken
from \cite{Rel11} and \cite{Smi2010}. }%
\label{fig:traps}
\end{center}
\end{figure*}

A challenge of hybrid atom-ion setups is to integrate overlapping
traps for ions and atoms into a single apparatus. Depending on
what kind of experiments are planned, the setups will look
somewhat different. Currently, all groups in the field use Paul
traps to confine the ions. However, the specifics of the design
and in particular the dimensions of the Paul traps depend on the
type of atom trap that is employed, in particular the magneto-optical trap
(MOTs)$\,$\cite{Rav12b,Sch12,Hal11}, the
far-off-resonant optical trap or the magnetic
trap$\,$\cite{Smi12,Zip11,Haz13} (see Fig.$\,$\ref{fig:traps}).

When working with a MOT, relatively large laser beams with
diameters of several mm are required. Thus, the Paul traps in such
setups feature large spacings between their radiofrequency (rf)
electrodes on the order of 1$\,$cm. Large Paul traps have the
advantage that they allow for loading
large ion crystals with hundreds or thousands of ions.
These ion crystals can then be overlapped with the MOT atom clouds
consisting of roughly $10^6$ atoms at temperatures around
100$\,\mu$K and densities of about $10^9\,$cm$^{-3}$. Such a setup
is well suited for investigations of chemical reactions such as
charge transfer or molecule formation in the cold regime.
Measurements can be carried out in a continuous way, monitoring
the fluorescence signals of the MOT and the trapped ions. However,
in a MOT a good fraction of the atoms are in an excited electronic state
which complicates the analysis of experimental signals.\\

For experiments where the atoms need to be in their electronic
ground state, at low atomic temperatures or at high atomic
densities, a MOT is not sufficient anymore. Instead, the atoms
need to be trapped either in a magnetic trap or a
far-off-resonance optical dipole trap. These atom traps allow for smaller 
 ion traps (e.g. on the order of 1$\,$mm) which yield
tighter ionic confinement. This comes at the price that the atomic
cloud has to be prepared elsewhere and subsequently transported to
the location of the ion trap. For example, in our atom-ion
apparatus in Ulm we first prepare an ultracold atomic cloud at a
temperature of 1$\,\mu$K in a magnetic trap using evaporative
cooling. Afterwards, we transport the atoms over a distance of
30$\,$cm to the ion trap, using a moving optical
trap$\,$\cite{Smi12}.

Working with a single ion (as compared to Coulomb crystals
consisting of many ions) generally improves the control over the
ion, e.g. in terms of its position, kinetic energy and internal
state. On the downside, using single ions can be time-consuming
when rare events need to be investigated with significant
statistics.

Ion traps and optical dipole traps for neutral atoms are in
principle quite compatible with each other. The inhomogeneous
electric fields $\varepsilon_\text{trap}$ of the ion trap only
lead to a relatively weak deconfining anti-trapping potential
$\propto 0.5 \alpha \varepsilon_\text{trap}^2$ for neutral atoms
with a static polarizability $\alpha$. For
Rb in our setup the corresponding ``anti-trap'' frequency is on
the order of a few Hz. This is much smaller than the trap
frequency of a typical optical dipole trap which ranges around one
hundred Hz. Similarly, the influence of the optical trap on the
ions is small compared to the deep Paul trap with its typical
trapping frequencies on the order of 100$\,$kHz.

\begin{figure}
\begin{center}
\resizebox*{8cm}{!}{\includegraphics{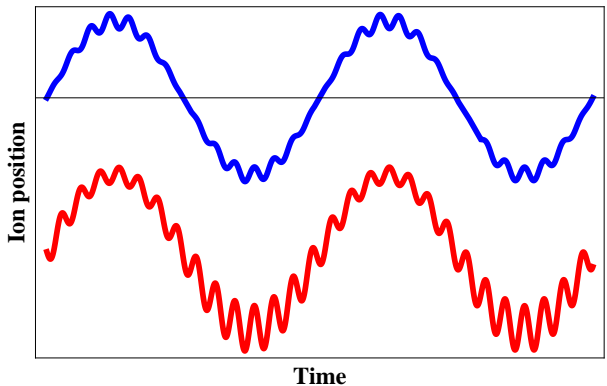}}%
\caption{Illustration of ion trajectories in a Paul trap (in the
absence of a neutral atomic cloud). \textit{Upper trace}: In an
ideal configuration, the ion oscillates symmetrically about the
nodal line of the rf field of the trap. Micromotion occurs
predominantly at the turning points of the secular oscillations
and vanishes when the ion passes through the rf node.
\textit{Lower trace}: An external electric field shifts the ion
from the center of the trap. The ion experiences larger rf fields and
thus undergoes excess micromotion oscillations. The energy
contained in the excess micromotion can
be much larger than the energy of the secular motion.}%
\label{fig:micromotion}
\end{center}
\end{figure}

An important experimental issue in a hybrid atom-ion setup is the
micromotion in the Paul trap$\,$\cite{Ber98}. This is linked to
the fact that, strictly speaking, the confining potential of a
Paul trap is not conservative as it is based on oscillating
electric fields. On top of the ion's so-called secular motion in
the effective harmonic trapping potential, the ion also carries
out a fast oscillatory motion (typically in the MHz range), the micromotion.
  The micromotion oscillations are a direct consequence of the 
  oscillating electrical forces of the rf field on the ion.
 Figure \ref{fig:micromotion} shows typical ion
trajectories in one dimension in a Paul trap. The upper blue trace
shows the motion of an ion oscillating symmetrically about the
trap center. As the rf fields vanish at the center and increase
when moving outwards, micromotion oscillations are strongest at
the turning points of the harmonic secular motion. Technical
imperfections of the ion trap can further enhance micromotion,
leading to the so-called excess micromotion. For example, a stray
electric offset field can shift the equilibrium position of the
trapped ion, as illustrated by the lower red trace of the ionic
trajectory. Furthermore, a phase delay of the rf voltage on
opposite Paul trap electrodes also leads to excess micromotion.
Here, the rf node, where the rf fields vanish, oscillates in
space. Excess micromotion can be minimized by compensating the
electrical stray fields and phase delays.

In current atom-ion experiments the typical energy scale for
excess micromotion is  on the order of 1$\,$mK$\times k_B$ or
above so that it sets the dominant energy scale. In an atom-ion
collision, the ionic micromotion energy can be transferred to the
atom. The energy that the ion loses in such a collision will then
quickly be replenished by the rf drive of the Paul trap. Thus, one
effect of micromotion is that it leads to ion-induced heating of
the neutral atomic cloud. It can also lead to atom loss if the
atom-ion collision is strong enough that the atom leaves its trap.

\section{Elastic atom-ion collisions }

Collisions are elastic when the internal state of the colliding
particles is not changed. As discussed in section 2, the elastic
cross sections for atom-ion collisions are typically orders of
magnitude larger than for cold neutral atoms due to the longer
range of the 1/r$^4$ potential. Such large cross sections are
useful to reduce time constants for particle interactions. For
example, possible applications are sympathetic cooling of ions
with a cold atomic gas or fast atom-ion quantum gates.
Furthermore, as we will discuss in section 6, strong atom-ion
interactions are of interest in experiments with polaronic
impurities aiming at the regime of strong coupling.

So far, experiments on elastic collisions between ions and cold
atoms have been carried out with combinations of Yb$^+$, Ba$^+$,
Rb$^+$, Ca$^+$, and Rb, Li (see e.g. \cite{Zip10, Smi2010, Rav12,
Har13, Haz13}). Typically, in these measurements one or more
trapped ions are immersed into a trapped gas of about 10$^5$ cold
atoms. Elastic collisions between an ion and atoms lead to: 1)
heating of the atomic cloud, 2) atom losses of the cloud, and 3) cooling
or even heating of the ion. The collision rate $\Gamma_e$ is given
by
\begin{equation}
\Gamma_e= n_\text{at} \sigma_e  v_\text{ion}\, ,
\label{eq:scattrate}
\end{equation}
where $v_\text{ion}$ is the velocity at which the ion moves
through the atom cloud and $n_\text{at}$ is the local atomic density. From
measurements of the decay of the atom number and the evolution of the
atomic temperature one can obtain information on the cross
sections and the kinetics of the ions.

A large ensemble of ultracold atoms at nK temperatures constitutes
a buffer gas that can be employed as a coolant for ionic collision
partners. Indeed, various experiments have demonstrated this
atom-ion sympathetic cooling. In these experiments the kinetic
energy of the single ion needs to be measured. Currently, the most
precise methods for doing this are based on
resolved-sideband cooling techniques$\,$\cite{Die89}. However, as
these techniques are quite involved, they have not yet been
applied to measure ion energies following atom-ion collisions.
Rather, a simpler but less precise temperature measurement, termed
ion recooling, was used to show cooling of ions in an ultracold
gas from much hotter temperatures to below
1$\,$K$\,$\cite{Zip10}(see also \cite{Rav12}). Ion recooling
measures the time it takes to laser cool an ion to Doppler
temperatures while observing its fluorescence. Yet another (but
more indirect) method to measure ion temperatures involves
atom-ion collisions. Given the cross section and the atomic
density one can infer the ion velocity from observed collision
rates. In an experiment with a single $^{87}$Rb$^+$ ion and cold
$^{87}$Rb atoms, sympathetic cooling from several Kelvin down into
the sub-mK regime
 was demonstrated$\,$\cite{Har12,Har13b}. This result seems promising
for a wide applicability of ultracold buffer gas cooling, as
$^{87}$Rb$^+$ is not amenable to laser cooling.

The question is how far down (beyond the mK-regime) one can
sympathetically cool ions using ultracold atoms in a hybrid trap.
Let us consider a single trapped ion which initially has been
laser cooled to Doppler temperatures of about 1$\,$mK. This ion is
immersed into a cold atomic cloud with a temperature below
1$\,\mu$K. Ignoring micromotion for a moment, one would expect the
ion to be sympathetically cooled down to the temperatures of the
atoms. Micromotion, however, prevents the ion from coming to rest
but keeps it at kinetic energies that are set by the excess
micromotion. One key to low ion energies is thus to minimize
excess micromotion. This can be done with the help of atoms as we
will discuss next.

\begin{figure}
\begin{center}
\resizebox*{8cm}{!}{\includegraphics{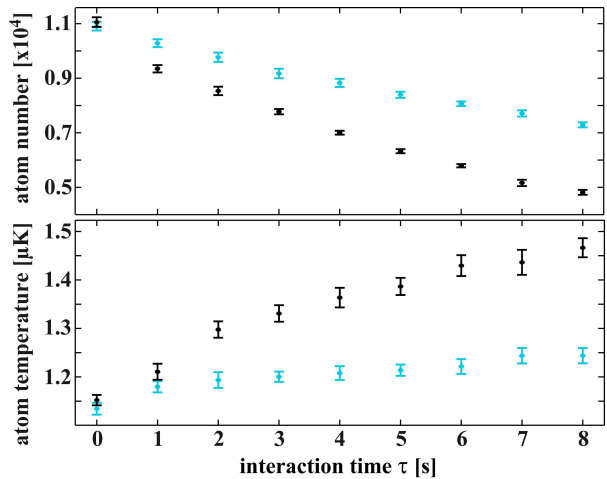}}%
\caption{As a consequence of the driven ion micromotion,
a Rb ion induces continuous atom loss (upper panel) and
heating (lower panel) in an ultracold Rb
atom cloud. The measurement was performed for
electric bias fields compensated to better than
$\varepsilon<0.1\,$V/m (blue data points) and a deliberately
applied offset field of about 4$\,$V/m (black data points).
Figure taken from \cite{Har13}.}%
\label{fig:atomloss}
\end{center}
\end{figure}

\begin{figure}
\begin{center}
\resizebox*{8cm}{!}{\includegraphics{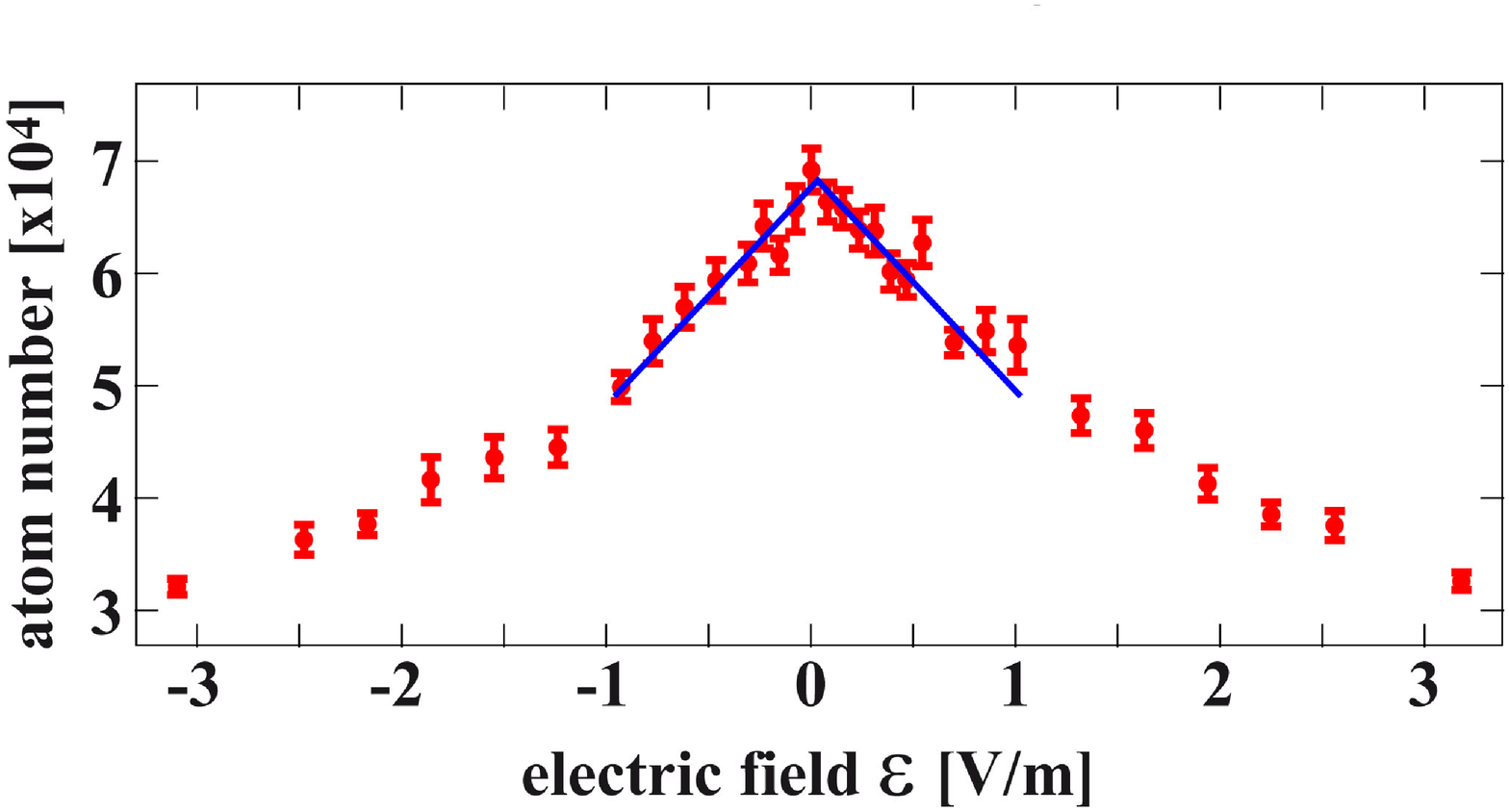}}%
\caption{ Micromotion compensation. Shown is the remaining atom
number of a cold atom cloud after an $8\,$s long interaction with
a single ion for various electrical offset fields $\varepsilon$. A
small $| \varepsilon |$ leads to small micromotion, minimizing
atomic losses.
Figure adapted from \cite{Har13}.}%
\label{fig:komp_dicht}
\end{center}
\end{figure}

On average, in a collision with an atom, the ion pumps energy into
the atomic cloud, heating it up and leading to atom loss. This has
been observed, e.g. in $\,$\cite{Har13} (see
Fig.$\,$\ref{fig:atomloss}). Here, a single Rb$^+$ ion is placed
at the center of a Rb atom cloud. At a density of about
$n_\text{at}=10^{11}\,$cm$^{-3}$ a continuous atom loss
and heating effect on timescales of several seconds are observed.
When excess micromotion is reduced (by compensating the stray
electric fields $\varepsilon$), the mean kinetic energy of the ion
decreases. This lowers heating and losses of the atomic cloud. It
is instructive to consider the micromotion energies $E_\text{eMM}$
involved in this measurement. If we ignore the contribution to
micromotion due to phase differences we can calculate the ion
excess micromotion energy due to stray electric
fields$\,$\cite{Ber98}. For the well-compensated case
($\varepsilon<0.1\,$V/m) we get $E_\text{eMM}=10\,\mu$K$\times
k_\text{B}$ while in the electric field of 4$\,$V/m we get
$E_\text{eMM}=20\,$mK$\times k_\text{B}$, an increase by more than
three orders of magnitude.
%
The sensitivity of the atomic loss on the ion energy can be used
to directly minimize micromotion. Figure$\,$\ref{fig:komp_dicht}
shows the remaining atom number as a function of the electric
offset field after an ion has been immersed for $\tau=8\,$s. The
peak in the atom number indicates the best compensation setting.
By performing a simple fit to the data, the electric field can be
minimized to better than 0.02$\,$V/m, corresponding to excess
micromotion energies (due to stray electric fields) below
1$\,\mu$K$\times k_\text{B}$. This result is comparable to the
most precise micromotion compensation methods based on ion
fluorescence
signals$\,$\cite{Har13,Ber98,Raa00,Nar11,Tan12,Chu13}.

It needs to be pointed out that even when micromotion due to both
stray electric fields and phase difference is compensated down to
the $\mu$K level the mean kinetic energy of the ion can still be
much larger. Besides anomalous heating
effects$\,$\cite{Tur00,Des06} where electrical field fluctuations
from the trap electrodes heat up the ion, collisional effects
between atoms and ions also lead to heating. This is discussed in
the following two sections.

A more refined theoretical analysis$\,$\cite{Maj68,DeV09,Zip10b}
of the atom-ion dynamics in a trap shows that the energy
distribution of the ion is non-thermal as it features a high
energy tail. The prominence of the high energy tail depends
critically on the atom-ion mass ratio. For a light ion interacting
with much heavier atoms, the collision dynamics can even heat the
ion to much higher ``temperatures'' than the one of the trapped
atoms. This counterintuitive effect is due to a transfer of energy
from micromotion into secular motion occurring in
atom-ion collisions.\\
However, it turns out that as long as the ionic mass is larger
than the mass of the neutral atom, the heating effect is moderate
and the energy distribution is close to thermal. The average ionic
energy can be then approximated by the plain excess micromotion
energy (i.e. in the absence of the neutral atom gas) multiplied by
a constant factor of order unity$\,$\cite{Zip10b}.\\
One further possible micromotion-induced heating mechanism is
discussed in ref.$\,$\cite{Cet12}. In this case, ion excess
micromotion is assumed to be perfectly compensated and the ion is
initially at rest at the trap center. When this ion is approached
by a neutral atom, the atom-ion interaction pulls the ion away
from the trap center into regions of non-vanishing micromotion.
Consequently, during the collision, the rf drive of the trap can
pump energy into the system. The average amount of injected energy
and thus the limit for the achievable collision energies depend on
the trap parameters and on the atom-ion mass ratio. To minimize
the heating effect a heavy ion (such as Yb$^+$) in a gas of light
atoms (such as Li) should be chosen.

Ultimately, micromotion-induced heating may set an effective limit
on the collision energies that can be realistically achieved in a
Paul trap environment. One viable path to circumvent such
limitations may be the use of an optical dipole trap to trap ions.
However, such trapping schemes are challenging as already small
electric stray fields on the order of 0.1$\,$V/m exert
comparatively strong forces on the ion which are hard to overcome
with optical dipole traps. Despite these difficulties,  a
proof-of-principle experiment for optical trapping of an ion has
been successfully carried out $\,$\cite{Sch10b}. A Mg$^+$ ion,
initially trapped in a linear Paul trap, was transferred into an
optical dipole trap consisting of 190$\,$mW of tightly focussed
laser light (beam waist $\approx 7 \mu$m) at a wavelength of
280$\,$nm. The achieved trap depth was about 40$\,$mK$\times
k_\text{B}$ and the maximal confining force was on the order of
$10^{-19}\,$N.  Spontaneous scattering of light led to strong
heating which limited the ion trap lifetime to about 2$\,$ms.

\section{Inelastic collisions and cold chemical reactions}
\label{sec:chargetransfer}

We now turn to inelastic collisions and reactions. We will first
consider charge transfer where the charge of the ion ``hops'' onto
the neutral collision partner. Next we discuss quenching
collisions where internal states of atomic or molecular ions relax
to energetically lower lying states via collisions with neutral
atoms. Finally, we describe some first observations of cold
atom-ion reactions in which molecules are formed.

\subsection{Charge exchange reactions}
\label{sec:charge_exchange}

\begin{figure*}
\begin{center}
\resizebox*{15cm}{!}{\includegraphics{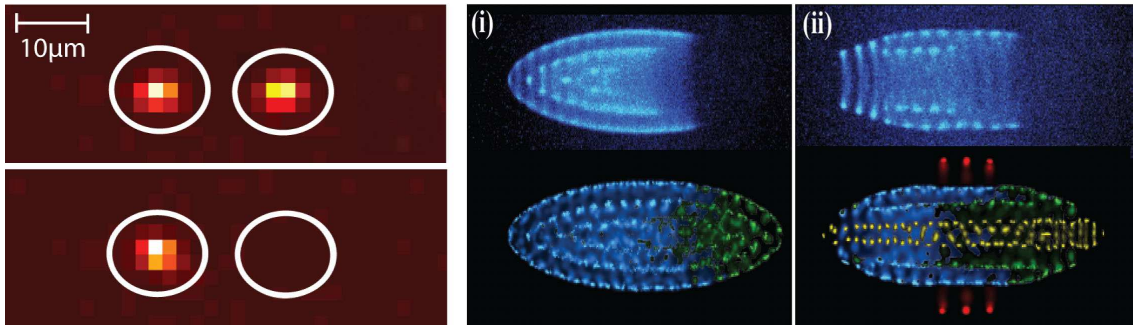}}%
\caption{Experimental signals of chemical reactions in hybrid
traps. \textit{Left}: Observing a charge exchange reaction in a string of two ions
in a linear Paul trap. Initially, two $^{138}$Ba$^+$ ions are laser cooled
and then brought into contact with a $^{87}$Rb atom cloud.
When one ion turns dark, this is indication that a reaction has taken place.
With subsequent measurements the mass of this dark ion can then be determined.
\textit{Right}: Charge exchange reactions using a large ion crystal with up to
several hundred ions. (i) Experimental
picture (upper image) and simulation (lower image) of a Ba$^+$ ion
crystal before interaction. (ii) Images after interaction with
$^{87}$Rb atoms. In the simulation, $^{138}$Ba ions are shown in
blue, lighter Ba$^+$ isotopes in green, $^{87}$Rb$^+$ ions in
yellow and BaRb$^+$ molecular ions
in red. By comparing the experimental pictures to the simulations, the product
fractions after the reaction can be inferred.
Figures taken from \cite{Smi2010} and \cite{Hal11}}.%
\label{fig:chargetransfer}
\end{center}
\end{figure*}

In the charge exchange reactions described here, a trapped ion
$A^+$ collides with an ultracold atom $B$ to produce $ A + B^+$.
In this process, particle $A$ is lost and replaced by $B^+$ which
is now trapped in the deep ion trap. Such a process can be
detected on a single particle level, e.g. via the loss of ion
fluorescence signals$\,$\cite{Gri09,Smi2010,Zip10} (see
Fig.$\,$\ref{fig:chargetransfer}). For example, in the experiments
in Ulm, the fluorescence of a single laser-cooled Ba$^+$ ion is monitored.
As soon as this ion undergoes a charge exchange with neutral Rb, the
ion is turned into a Ba atom and it stops fluorescing. The created
Rb$^+$ ion does not scatter light in the visible and is ``dark''. 
Such a ``dark'' ion can be investigated via mass spectrometry in the
Paul trap.  This type of investigation can be done with a single ion or with
many ions that form a crystal of ``dark''
and ``bright'' ions (see Fig.$\,$\ref{fig:chargetransfer}).
 When the drive frequency of the trap is
modulated at a secular frequency of a ``dark'' ion, this ion is
strongly heated. Via Coulomb coupling between the ``dark'' and ``bright'' ions, 
 the ``bright'' ions are
sympathetically heated and their fluorescence signals become
blurred or are even lost completely. This method is routinely used
to analyze the contents of mixed ion crystals consisting of up to
thousands of ions$\,$\cite{Rel11,Hal11} (see
Fig.$\,$\ref{fig:chargetransfer}).

\begin{figure}
\begin{center}
\resizebox{8cm}{!}{\includegraphics{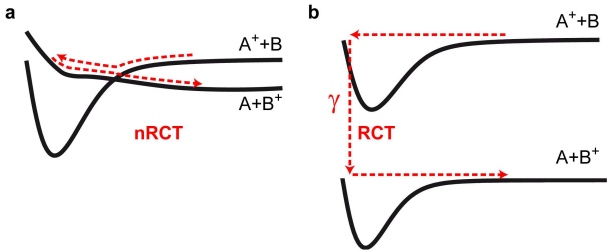}}%
\caption{Charge transfer. a) An avoided crossing of molecular
potential curves can favor non-radiative charge transfer (nRCT) if
the crossing is not fully adiabatic or non-adiabatic.   b)
Radiative charge transfer involves the emission of a photon
$\gamma$.
  }%
\label{fig:chargetransfersketch}
\end{center}
\end{figure}

Charge transfer can proceed in two different ways: (1) through
emission of a photon $\gamma$ in a radiative charge transfer (RCT)
$A^++B\rightarrow A+B^++\gamma$, or (2) in a non-radiative charge
transfer (nRCT), $A^++B\rightarrow A+B^+$. The probability for
these charge exchange reactions to occur in a collision can vary
by several orders of magnitude. It depends on the details of the
colliding particles and their electronic states. For example, the
probability for nRCT can be very large for the case of a crossing
of energy levels of the two scattering channels $A^++B$ and
$A+B^+$ (see Fig.$\,$\ref{fig:chargetransfersketch}a).

Already a good number of charge exchange processes in cold
atom-ion collisions have been studied, involving collisions of ions
such as Yb$^+$, Ba$^+$, Ca$^+$, N$_2^+ $ with neutral atoms such as Yb,
Rb and
Ca$\,$\cite{Gri09, Smi2010, Zip10a, Rat12, Rel11, Hal11, Hal12, Sul12, Hal13b}.
In general, these experiments confirmed that the charge
transfer rate coefficients are bounded from above by the classical
Langevin collision cross section (see  section
\ref{sec:potential}). This seems reasonable when we assume that
charge transfer requires a close encounter of ion and atom, i.e. a
Langevin collision. Making use of its cross section
we can calculate the corresponding collision rate
\begin{equation}
\Gamma_\text{Lgvn}= n_\text{at} \sigma_\text{Lgvn} v_\text{ion},
\end{equation}
where the ionic velocity $v_\text{ion}\propto E_\text{col}^{1/2}$
is assumed to be much larger than the atomic velocity. With
$\sigma_\text{Lgvn}\propto E_\text{col}^{-1/2}$ (see
equation$\,$\ref{eq:ct}) we find that $\Gamma_\text{Lgvn}$ should
be independent of the collision energy, a fact which was confirmed
experimentally \cite{Gri09,Zip10a,Rel11,Hal13a,Hal13b}. While
bounded from above, charge exchange rates can be very much smaller
than the Langevin collision rate. Suppression factors ranging from
about 1 to $10^{-5}$ have been measured. The magnitude of the
suppression factor is determined by the details and the mechanism
of the collision, i.e. whether it involves emission of a photon
and what the precise form of the relevant molecular potential
energy curves is.

An exception to the applicability of Langevin theory was observed
in investigations of reactions between N$_2^+$ molecular ions with
Rb atoms in the excited $^2P_ {3/2}$ state$\,$\cite{Hal12}. In
these measurements, the charge exchange rate was found to be 4
times larger than the maximal value allowed within Langevin
theory. This large rate can be explained by taking into account
the permanent electric quadrupole moment of the excited state of
neutral Rb that interacts with the ion. This interaction adds to
the polarization potential considered in Langevin theory and
further increases the reaction cross section.

While charge exchange reactions provide an interesting field to explore,
they are often undesired in atom-ion experiments.
In order to avoid charge exchange completely, one can choose appropriate
combinations of species (such as Ba$^+$ and Ca$\,$\cite{Sul12,Rel13}) in
which chemical reactions are prohibited energetically, at least in
ground state collisions.

\subsection{Quenching reactions}
\label{sec:quenching}

In addition to charge exchange, collisions of atomic (or
molecular) ions with neutral atoms can lead to a relaxation of
internal excitations in so-called quenching reactions.
Quenching can be applied to sympathetically cool internal
excitations in addition to the external motion (as previously
discussed in section$\,$4).\\
In ref.$\,$\cite{Rat12}, Yb$^+$ ions were prepared in
electronically excited metastable states and immersed into
ultracold Rb atom clouds. Quenching in these
experiments was detected by shining in resonant laser light,
leading to ionic fluorescence signals as soon as the quench from
the metastable states into the
ground state $^2S_{1/2}$ occurred. While the excitation into the
$^2F_{7/2}$ state of Yb$^+$ was apparently efficiently quenched,
no such reaction was observed upon preparation of the $^2D_{3/2}$ state.
It will be interesting to understand the physical origin of these
different quenching dynamics in future investigations.\\
In experiments with molecular BaCl$^+$ ions immersed into a cloud of
ultracold Ca atoms, quenching of internal vibrational excitations
of the molecular ions was achieved$\,$\cite{Rel13}. This
particular combination of atomic and ionic species was chosen
because of the stability with respect to charge exchange
reactions.
Clouds of BaCl$^+$ ions were prepared in a Paul trap and their
internal degrees of freedom equilibrated to the temperature of the
surrounding vacuum chamber which was at room temperature. In this
initial setting, the vibrational ground state of the molecules was
populated with a probability of about 80$\,$\%. After immersion of
these molecules into an ultracold cloud of Ca atoms produced in a
MOT, the ground state population increased to 90$\,$\%, only limited by
heating effects induced by black-body radiation from the vacuum
chamber walls. The quenching rate constant
$k_q=1\times10^{-9}\,$cm$^{3}$s$^{-1}$ was found to be large and
it indicates that every
fifth Langevin collision leads to a quench of the vibrational excitation.\\

\subsection{Formation of charged molecules and molecular reactions}
\label{sec:interstellarchemistry}

The controlled formation of cold ionic molecules is an important
goal in the field of atom-ion experiments. It is related to
research of chemistry in interstellar clouds where molecules are
formed under dilute (density $<10^6\:\text{cm}^{-3}$) and cold
(temperature $\sim 10\:$K) conditions$\,$\cite{Smi92,Ger92}. A
large number of groups have contributed to this field with
beautiful experiments with collision energies down to the low
K$\times k_B$ regime (see e.g. \cite{Rot06, Sta08, Wil08, Dei12}).
Here, however, we will constrain ourselves to laboratory experiments
in the mK regime. In this regime, to date only a few chemical reactions
have been observed.\\
In reference$\,$\cite{Hal11,Hal13b} the collision of Ca$^+$ and Rb
was studied.  It was found that fast reactions occur when the
Ca$^+$ ions are excited to their $4p$ state. As reaction products,
both Rb$^+$ due to charge transfer and RbCa$^+$ molecular ions
were detected. These molecular ions were probably produced via
radiative association, a process in which a photon is emitted upon
the atom-ion collision and a molecular ion is formed:
$A^++B\rightarrow (AB)^++\gamma$. Both the atomic and the
molecular ions in these experiments were observed within a large
ion crystal in the Paul trap. They were identified using the mass
spectrometry methods discussed in
section$\,$\ref{sec:charge_exchange}.

\begin{figure}
\begin{center}
\resizebox*{8cm}{!}{\includegraphics{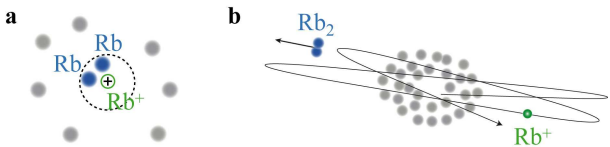}}%
\caption{Illustration of the effect of an atom-ion three-body collision. The large amount of released binding energy
ejects the ion from the atom cloud and promotes it onto a large orbit within the ion trap. This effectively stops the atom-ion
interaction for long time periods on the order of seconds.}%
\label{fig:cartoon}
\end{center}
\end{figure}

Another pathway to the formation of molecular ions is three-body
recombination, where the ion and an atom combine to form a
molecule and a second atom carries away part of the binding
energy. In section$\,$\ref{sec:potential} we discussed the large
atom-ion interaction radius $R^*$ on the order of hundreds of
nanometers. As a consequence, there is a significant probability
for two atoms to enter the atom-ion interaction range
simultaneously, even at moderate atomic densities of
10$^{12}$cm$^{-3}$. Collisions involving a single ion and two
neutral atoms thus become relevant at much lower densities than in
the well-known case of interatomic three-body collisions. In
neutral atomic gases three-body recombination typically leads to
the loss of all three atoms involved in the
reaction$\,$\cite{Hess1983,Burt1997,Esry1999} as the released
binding energy kicks the particles out of trap. In the case of
atom-ion three-body recombination, the ionic product will
typically not be lost as it is retained in the very deep Paul
trap. In an experiment with a Rb$^+$ ion in a cold Rb cloud
\cite{Har12} the energy released in a three-body recombination
event was found to be on the order of a few 0.1$\,$eV, much
smaller than the ion trap depth. Although still trapped, the ion
was ejected out of the 100 $\mu$m-sized atom cloud upon the
recombination (see Fig.$\,$\ref{fig:cartoon}). The ion's
trajectory in the trap was so large that it barely interacted with
the atomic sample at the trap center. Thus, for time periods of a
few seconds, the atom-ion interaction effectively stopped, leading
to a drastic reduction of atomic loss. Interestingly, the produced
molecules could not be detected directly - at the end of each
experiment exclusively an atomic Rb$^+$ ion was observed. As a
consequence it could not be determined whether Rb$_2$ or Rb$_2^+$ molecules
had been initially produced, as the Rb$_2^+$ molecules might have
subsequently quickly disintegrated back to Rb$^+$ in another light
assisted collision with Rb atoms.

The production of Rb$_2^+$ molecules was directly observed in
another experiment \cite{Har13c}. The production path, however,
starts with neutral atoms. A relatively dense atomic cloud of
neutral Rb atoms (10$^{13}$cm$^{-3}$) was placed inside a Paul
trap. Via interatomic three-body recombination in this cloud,
Rb$_2$ molecules were formed. Some of these neutral molecules were
then resonantly photoionized and subsequently trapped in the Paul
trap where they were detected. Interestingly, when bringing these
molecular Rb$_2^+$ ions back into contact with the optically
trapped Rb gas they quickly dissociate (on the ms timescale) into
atomic Rb$^+$ ions, as discussed in the previous paragraph.
The dissociation mechanism is not yet fully
understood and will be the subject of further investigations.

\section{Future prospects for atom-ion experiments}

Hybrid systems of ions and atoms also lend themselves to
experiments on topics of quantum information science and condensed
matter physics. In the following we briefly discuss some
theoretical schemes that have been proposed in these contexts.

\subsection{Atom-ion Feshbach resonances}
\label{sec:feshbach}

\begin{figure}
\begin{center}
\resizebox*{7cm}{!}{\includegraphics{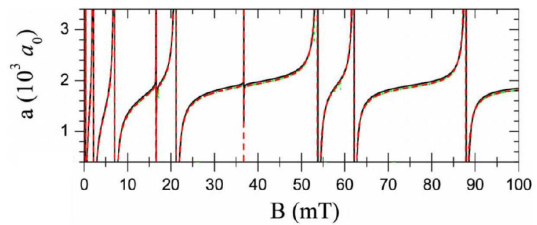}}%
\caption{Calculated Feshbach resonances for Ca$^+$ colliding with
Na. Shown is the scattering length $a$ versus the magnetic field
$B$. Already at moderate magnetic fields below 10$\,$mT a number of
resonances are predicted. Figure adapted from$\,$\cite{Idz09}.}%
\label{fig:feshbach}
\end{center}
\end{figure}

A key to the engineering of complex many-particle systems is the control over
the interparticle interactions. Such control is most easily
obtained in the regime of s-wave scattering where the interaction
can be described by a single quantity, the
scattering length. Similar to established techniques in neutral
atomic gases, atom-ion Feshbach resonances could be used to
tune the scattering length. The atom-ion
interaction could thus be set strongly attractive to
non-interacting to repulsive. Furthermore, Feshbach resonances
would be convenient tools to efficiently form charged molecules in
precisely defined quantum states. Indeed, such resonances have
been predicted to exist in collisions of singly ionized alkaline
earth atoms with alkali atoms, such as Ca$^+$ + Na \cite{Idz09}
(see Fig.$\,$\ref{fig:feshbach}).
Interestingly, resonances should appear already at relatively low
magnetic fields on the order of tens of Gauss (1$\,$G = $10^{-4}$T).
This can be explained as a consequence of the relatively
long-range 1/r$^4$ potential which leads to many weakly bound
molecular states. In order to make use of Feshbach resonances, the s-wave regime
needs to be reached, which typically corresponds to 
collision energies of 100 nK- 1 $\mu$K$\times k_B$. This is a challenge.

\subsection{Prospects for quantum information applications}
\label{sec:decoherence}

The great success of both ions and atoms in the context of quantum
information science obviously raises hopes that a combined
atom-ion system may provide additional opportunities for this
field. Ions, on the one hand, are routinely controlled on the
single particle level and their tight traps allow for
fast and efficient transport schemes. Neutral atoms can exhibit
long coherence times and large arrays of atoms can be prepared in
an optical lattice with a single atom per lattice site,
constituting a quantum register. An ion may then be used to
transport quantum information between different sites of the
optical lattice and to entangle atoms. Similarly, a single ion can also
read out information  so that it can act as a read and write head. 
The transfer of quantum
information between the ion and the atoms could be achieved via
controlled collisions where both the ion and the
atom are held in their individual trap potentials$\,$\cite{Idz07}. A special
situation arises here due to the long-range atom-ion interaction
potential in combination with the tight trapping potentials.
Depending on the distance between atom and ion, this can lead to
trap-induced shape resonances or confinement-induced resonances.
In both cases, vibrational trap states become equal in energy to
molecular states giving rise to avoided crossings. It has been
shown that this can be exploited to implement atom-ion quantum
gates$\,$\cite{Doe10}. However, it must be said that ion
micromotion severely complicates the atom-ion collisional
processes and could make the experimental realization of the
proposed schemes challenging$\,$\cite{Ngu12}.

\begin{figure}
\begin{center}
\resizebox*{8cm}{!}{\includegraphics{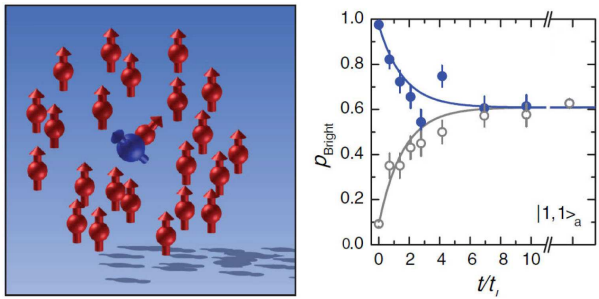}}%
\caption{A single Yb$^+$ ion as an impurity in a spin-polarized
atomic gas. The $^{174}$Yb$^+$ qubit is initially prepared in the
state $m_J = 1/2$ (full symbols) or $m_J = -1/2$ (open symbols).
Within a few Langevin collision time periods $t_L$ in a gas of Rb
atoms ($|F=1,m_\textrm{F}=1\rangle$) the state relaxes to a mix of
$m_J = \pm 1/2$. $P_{\text{Bright}}$ is the probability of the ion
to occupy the state $m_J = 1/2$. Figure adapted from \cite{Rat13}. }%
\label{fig:spinbath}
\end{center}
\end{figure}

In ref.$\,$\cite{Ger12} an ion is proposed to act as a quantum
switch for the transport of neutral atoms between lattice sites.
It is conjectured that the switch can be made state dependent
which can be used to create controlled entanglement of atomic and
ionic
states.\\
Experimentally, the degree of control over atom-ion systems to
realize such quantum information schemes has not been reached yet.
However, first steps in this direction have been taken. For
example, the decoherence/ lifetime of an ionic qubit in an atomic
bath has been studied$\,$\cite{Rat13}. Here, an Yb$^+$ ion is prepared
as a single spin impurity in a bath of spin-polarized $^{87}$Rb
atoms (see Fig.$\,$\ref{fig:spinbath}). Resonant laser light is
used to control the population of the two Zeeman sublevels $m_J =
\pm 1/2$ of the ground state of $^{174}$Yb$^+$. In this way, a
qubit is realized and immersed into Rb atom clouds prepared in the
$|F=1,m_\textrm{F}=1\rangle$ ground state.
These first results show a very short coherence lifetime of the
qubit. After a few Langevin collisions the ionic state typically
relaxes and depolarizes in a spin-nonconserving way. This
relaxation mechanism might be due to a very strong spin-orbit
coupling which is typical for heavy elements like Yb$^+$ and Rb.
It will be interesting to see whether such strong spin relaxation can
also be seen in other atom-ion systems. From the standpoint of
quantum information applications, a system with negligible spin
relaxation but efficient spin exchange would be desirable.

\subsection{Polarons in the strong coupling regime }
\label{sec:polarons}

When immersing an immobile ionic impurity into a quantum gas, it
is expected that atom-ion interaction leads to a local modulation
in the quantum gas density \cite{Mas05,Goo10}.
This can be understood in terms of a dressing of the ion with the neutral
atoms. If the ion is mobile this dressing gives rise to a
quasi-particle, a polaron (see e.g. ref.$\,$\cite{Ale09}).
Ultracold gases have recently received considerable attention in
the context of polaron physics due to the excellent tunability of
the relevant parameters, such as the scattering length and the
particle density $\,$\cite{Cuc06,Kal06,Tem09,Cas11,Sch09,Nas09,Koh12,Kos12}.

Of special interest is polaron research in the strong coupling
limit where unsolved problems are still present \cite{Cas11b}. So
far, this limit has been elusive both in condensed matter systems
and in systems of cold neutral atoms. However, an ionic impurity
with its large interaction range is expected to easily enter the
strong coupling regime$\,$\cite{Cas11b}. Predicted phenomena in
this regime include self-trapping and collapse of
the condensate. Self-trapping is an especially intriguing
phenomenon where the impurity first induces a bubble in the
condensate as a consequence of its strong interaction. The bubble
retroactively leads to confinement of the impurity.
\\
\subsection{Mesoscopic molecular ions}
\label{sec:mesoscope}

The atom-ion polarons are related to the work in
reference$\,$\cite{Cot02} where a novel mesoscopic ionic molecule
is predicted. This molecule consists of a single ion at rest which
loosely binds many neutral atoms into the most weakly bound
molecular states of the 1/r$^4$ polarization potential. When
placed in a BEC, large shells form around the ion, which are
filled with up to hundreds of atoms. The shell size can reach
0.1$\,\mu$m and the binding energy is on the order of 0.1 to
1$\mu$K$\times$k$_B$. For Na, these mesoscopic states are
predicted to form spontaneously on the ms timescale (at densities
of 10$^{14}\,$cm$^{-3}$) after a Na$^+$ ion has been immersed into
a Na BEC. The released binding energy is either dissipated by
phonon emission in the condensate or by atom-ion three-body
collisions. The bound states can in principle decay into more
deeply bound states but lifetimes are predicted to be relatively
long, on the order of tens of milliseconds. This should make it
possible to experimentally investigate some of the properties of
these exotic mesoscopic quantum objects.

\subsection{ Coupling of crystal phonons to a Fermi gas}
\label{sec:Fermi}

Another recent proposal suggests to combine a cold atomic Fermi
gas with a trapped linear ion crystal$\,$\cite{Ger13}. The
crystal of trapped ions is regarded as an experimentally tunable
analog of a solid state lattice. The ultracold gas of fermionic
atoms, on the other hand, represents the electron gas of the solid
state lattice. The ion crystal and the atom cloud are brought into
contact or close to each other ($\approx 100\,$nm) to allow for
interaction. This interaction between atom gas and ion lattice
would then correspond to phonon-electron coupling in a solid state
phase. If disturbing effects of micromotion can be neglected, such
a combined system is potentially able to mimic interesting solid
state physics such as polaronic quasiparticles, Peierls
transitions, and the emergence of phonon-mediated interactions.
The tunability of the ultracold atomic gas combined with the high
fidelity operations and detection from trapped ions make this
system intriguing when comparing it to possibilities in the solid
state phase.

\subsection{Charge transport } \label{sec:chargetransport}

A further idea for cold atom-ion research concerns charge transport
mechanisms in an ultracold gas. In ref.$\,$\cite{Cot00b} a
transition in the charge transport mechanism is predicted in a
gas of Na atoms doped with ionic impurities, Na$^+$. At high
temperatures charge transport occurs via motion of ions.
Collisions of the ions with the gas limit their mobility. At low
temperatures of a few $\mu$K, however, charge transport
becomes dominated by hopping of electrons from neighboring atoms to the
ions.
This results in a fast diffusion of the charges through the
ultracold gas (of which the motion is then negligible) and the
system is expected to exhibit a drastic increase in charge mobility by several
orders of magnitude.

\section{Summary}
In conclusion, we have provided an introduction to the young field of cold hybrid
atom-ion experiments. After describing the fundamentals of the
1/r$^4$ interaction between an atom and an ion we have given a
brief overview of first experiments, ideas for future research as
well as challenges to be tackled. At this stage experiments in the
field are mostly focussed on elastic and reactive collisions
such as charge exchange, quenching processes or molecule
formation. These processes are still studied in an energy
regime where many partial waves contribute, a
fact that can be mainly attributed to residual micromotion of the
ion in the Paul trap. Indeed, many ideas for future research hinge
on reaching a regime where micromotion can be neglected. Proposals
range from applications in quantum information to condensed matter
related research. Obviously, as the field of cold
atom-ion experiments has just started to unfold, many more routes
for research can be envisioned. As an increasing
number of scientists are getting involved in this line of
research, we are looking forward to an accelerating development
and a growing number of exciting results to come.

\section*{Acknowledgements}
We would like to thank all members of our team for the nice
collaboration and discussions over the years. We are also grateful
to members of the COMIQ EU Marie Curie training network  and the
SFB/TRR21 for fruitful exchange. We would like to acknowledge
financial support from the German Research foundation DFG within
the SFB/TRR21.

\bibliographystyle{tCPH}

\end{document}